\documentclass[runningheads]{llncs}
\usepackage{booktabs}
\usepackage{graphicx}
\usepackage{multicol,listings,multirow}
\usepackage{lipsum}
\usepackage{tikz}
\usepackage{float}
\usepackage{amsmath}
\usepackage{makecell}
\usepackage{amssymb}
\usepackage{wrapfig}
\usepackage{array}
\usepackage{booktabs}
\usepackage{array}
\usepackage{color}
\usepackage[colorlinks=true,
            linkcolor=blue,
            filecolor=blue,      
            urlcolor=blue,       
            citecolor=blue
            ]{hyperref}
\usepackage[marginal]{footmisc}
\usepackage[misc]{ifsym}

\makeatletter 
  \newcommand\figcaption{\def\@captype{figure}\caption} 
  \newcommand\tabcaption{\def\@captype{table}\caption} 
\makeatother

\begin{document}
\title{BSDA-Net: A Boundary Shape and Distance Aware Joint Learning Framework for 
Segmenting and Classifying OCTA Images}

%

%
\author{Li Lin\inst{1,2}\orcidID{0000-0002-9789-0825}\and
Zhonghua Wang\inst{1} \and
Jiewei Wu\inst{1,2} \and
Yijin Huang\inst{1} \and
Junyan Lyu\inst{1} \and
Pujin Cheng\inst{1} \and
Jiong Wu\inst{3} \and
Xiaoying Tang$^{1(\textrm{\Letter)}}$}
%

\authorrunning{L. Lin et al.}
\titlerunning{BSDA-Net: Boundary Shape and Distance Aware Joint Learning Framework}
%
\institute{Department of Electrical and Electronic Engineering, Southern University of Science and Technology, Shenzhen, China\\ 
\email{tangxy@sustech.edu.cn}\and
SEIT, Sun Yat-Sen University, Guangzhou, China
\and
School of Computer and Electrical Engineering, Hunan University of
Arts and Science, Hunan, China}
\maketitle              
\footnote{L. Lin and Z. Wang contributed equally to this work.\\}
\begin{abstract}
Optical coherence tomography angiography (OCTA) is a novel non-invasive imaging technique that allows visualizations of vasculature and foveal avascular zone (FAZ) across retinal layers. 
Clinical researches suggest that the morphology and contour irregularity of FAZ are important biomarkers of various ocular pathologies. 
Therefore, precise segmentation of FAZ has great clinical interest.
Also, there is no existing research reporting that FAZ features can improve the performance of deep diagnostic classification networks. 
In this paper, we propose a novel multi-level boundary shape and distance aware joint learning framework, named BSDA-Net, for FAZ segmentation and diagnostic classification from OCTA images. 
Two auxiliary branches, namely boundary heatmap regression and signed distance map reconstruction branches, are constructed in addition to the segmentation branch to improve the segmentation performance, resulting in more accurate FAZ contours and fewer outliers.
Moreover, both low-level and high-level features from the aforementioned three branches, including shape, size, boundary, and signed directional distance map of FAZ, are fused hierarchically with features from the diagnostic classifier.
Through extensive experiments, the proposed BSDA-Net is found to yield state-of-the-art segmentation and classification results on the OCTA-500, OCTAGON, and FAZID datasets.

\end{abstract}
\section{Introduction}
Optical coherence tomography angiography (OCTA) is an emerging non-invasive ophthalmological imaging technique with an ability to generate high-resolution volumetric images of retinal vasculature. It has been increasingly recognized as an invaluable imaging technique to visualize retinal vessels and foveal avascular zone (FAZ) \cite{ma2020rose}.
OCTA \textit{en face} maps are produced by projection over
regions of selective depths, and can be basically divided into superficial images and deep ones with different fields of view (FOVs), e.g., 3 mm $\times$ 3 mm and 6 mm $\times$ 6 mm, obtained from different scan modes. The former mode has a higher scan resolution than the latter one, and thus the FAZ and capillaries are depicted more clearly. In contrast, the latter mode covers a broader area and has a greater ability in detecting pathological features such as microaneurysms and non-perfusion \cite{de2015review,leitgeb2019face}.
Many retinal biomarkers are extracted from the OCTA \textit{en face} maps (hereinafter collectively referred to as OCTA images), given that the flattened retinal structures in OCTA images are more informative and convenient for ophthalmologists to examine \cite{li2020image}.
Existing evidence suggests that the morphology and contour irregularity of FAZ are highly relevant to various ocular pathologies such as diabetic retinopathy (DR), age-related macular degeneration (AMD), and so on \cite{linderman2017assessing,salles2016optical,zheng2010automated}. For instance, patients with high myopia typically have reduced FAZ areas, whereas macular ischemia caused by diabetes have enlarged FAZ areas \cite{balaji2020comparison,ometto2020fast}. 
As such, precise FAZ segmentation is of great clinical significance. Moreover, OCTA as a new modality shows its potential in computer-aided eye disease and eye-related systemic disease diagnoses \cite{ma2020rose}.

In the past few years, several automatic FAZ segmentation algorithms have been proposed, which can be mainly divided into two categories. The first category is unsupervised methods, typically including statistical segmentation methods and mathematical morphology methods. For instance, Haddouche et al. \cite{haddouche2010detection} employed a Markov random fields based method to detect FAZ. Pipelines based on combinations of morphology processing methods and transformation methods also yielded reasonable results \cite{diaz2019automatic,lu2018evaluation,silva2015segmentation}. The past few years have witnessed a rapid development of the second category of methods, the deep learning based methods, which achieved overwhelming performance in almost all computer vision fields. Several works based on UNet \cite{ronneberger2015u} and its variants for FAZ segmentation have been reported \cite{guo2019automatic,li2020image,li2020ipn,li2020fast}. Despite their progress, these methods still have limitations, such as imprecise boundaries due to inferior image quality, confusing FAZ with interfering structures, generating inevitable outliers when there exists erroneous layer projection, and failing to segment low contrast FAZ from its surrounding region. Some representative images are shown on the left panel of Fig. \ref{fig1} and Fig. \ref{fig2}. 
The main reason for these problems is that UNet based methods typically lack the ability to learn sufficiently strong prior knowledge via single task/loss constraint on small medical image datasets. 
As for OCTA-based automatic diagnostic classification, only a few papers have reported their attempts. For example, Minhaj et al. designed several hand-crafted features such as vessel tortuosity and FAZ area, and employed support vector machine for classification and staging \cite{alam2019supervised}. Deep ImageNet-pretrained networks have also been proposed for classifying small-sampled OCTA datasets \cite{andreeva2020dr,le2020transfer}. However, these methods are still very preliminary given most of them only utilized pretrained models followed by fine-tuning OCTA images from the specific study of interest, and there is still room for improvement. Meanwhile, there is no research showing that explicit or implicit FAZ-related features can be effectively and interpretably utilized in deep classification networks.

In such context, we propose a novel hierarchical and multi-level boundary shape and distance aware joint learning framework for FAZ segmentation and eye-related disease classification utilizing OCTA images. 
Two auxiliary branches, namely boundary heatmap regression and signed distance map (SDM) reconstruction branches, are constructed in addition to the segmentation branch following a shared encoder to improve the segmentation performance.
Also, both low-level and high-level features from the aforementioned three branches, including shape, boundary, and signed directional distance map of FAZ, are fused hierarchically with features from the diagnostic classifier.

The main contributions of this paper are four-fold: (1) We present the first joint learning framework, named BSDA-Net, for FAZ segmentation and multi-disease (e.g. DR, AMD, diabetes and myopia) classification from OCTA images. (2) We propose boundary heatmap regression and SDM reconstruction auxiliary tasks, which can effectively improve the performance of FAZ segmentation in both joint learning and single-task learning settings. (3) Via hierarchically fusing features from decoders of the three segmentation-related tasks and the classifier, we demonstrate the effectiveness of FAZ features in guiding and boosting deep classification networks interpretably. (4) We validate the effectiveness of our method on three publicly-accessible OCTA datasets, and our approach achieves state-of-the-art (SOTA) performance in both tasks, establishing new baselines for the community. We make our code available at \url{https://github.com/llmir/MultitaskOCTA}.

\section{Methodology}
\begin{figure}[htbp]
    \centering
    \centerline{\includegraphics[width=\textwidth]{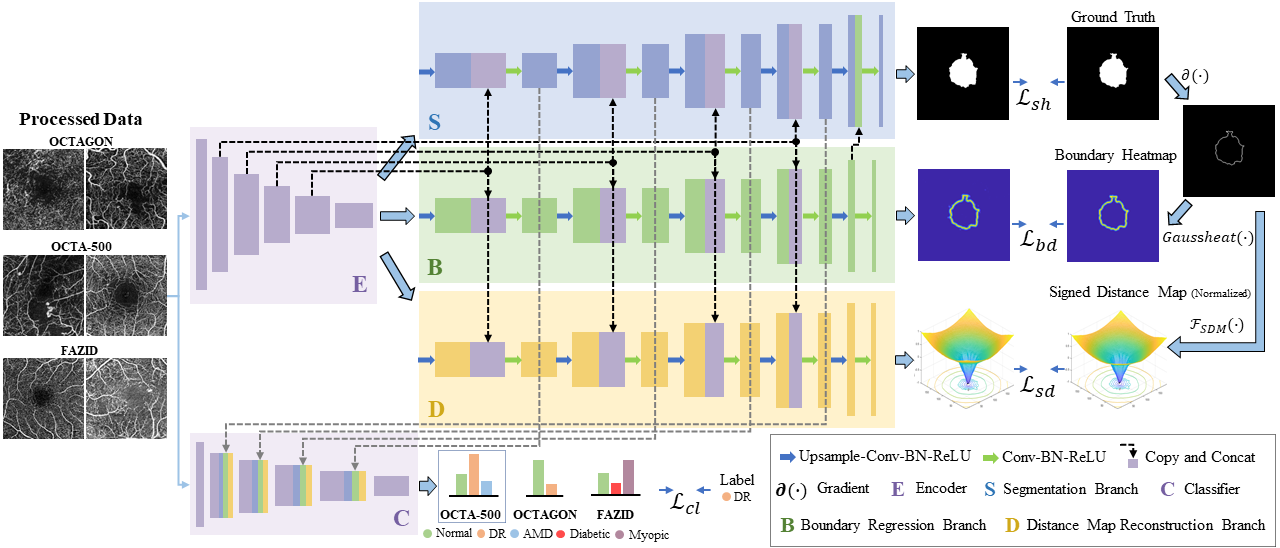}}
    \caption{Schematic representation of the architecture of our proposed framework.} \label{fig1}
\end{figure}

The proposed joint learning framework of BSDA-Net is shown in Fig. \ref{fig1}, which is composed of a multi-branch segmentation network (segmentor) and a classifier. Each component will be described detailedly in the following subsections.

\subsection{Segmentation Network}
As illustrated in Fig. \ref{fig1}, the segmentor in BSDA-Net adopts the widely employed encoder-decoder architecture, which is composed of a shared encoder $E$ and three different decoders, namely a segmentation branch $S$, a soft boundary regression branch $B$, and a SDM reconstruction branch $D$. 

BSDA-Net is a general framework, and any segmentation network with an encoder-decoder architecture, such as UNet \cite{ronneberger2015u}, UNet++ \cite{zhou2019unet++}, PSPNet \cite{zhao2017pyramid}, and DeepLabv3+ \cite{chen2018encoder}, can fit in. In this paper, we employ an adapted UNet structure with ResNeSt50 being the encoder for illustration \cite{zhang2020resnest}. The three decoders have the same structure, and features from the penultimate layer of branch $B$ are concatenated to features from the corresponding layer of branch $S$ for better contour perception and preservation. Each decoder module comprises nearest upsampling with a scale factor of 2, followed by two layers of 3 $\times$ 3 filters, batch normalization (BN), and ReLU. In our setting, three small decoders are adopted, with the initial number of feature maps being 256 and getting halved after every upsampling layer. 
Under small dataset conditions, the edge region of a predicted mask may be inaccurate and has a high probability of being over or under segmented. Also, interfering structures and erroneous layer projection in OCTA images typically lead to outliers.
We therefore propose a novel multitask combination by constructing two auxiliary tasks to reconstruct edges and SDM, which provide the encoder with more topological priors both explicitly and implicitly and make them collaborate with the primary segmentation task to obtain a more accurate target segmentation mask.


Given a target FAZ and a point $x$ in the OCTA image, the SDM of the ground truth $G$ is formulated as
\begin{equation}
G_{sd}=\mathcal{F}_{S D M}(\partial G)=\left\{\begin{array}{ll}
    -\inf _{y \in \partial G}\|x-y\|_{2}, & x \in G_{i n} \\
    0, & x \in \partial G \\
    \inf _{y \in \partial G}\|x-y\|_{2}, & x \in G_{o u t}
    \end{array}\right.,
\end{equation}
where $\|x-y\|_{2}$, $\partial G$, $G_{i n}$, and $G_{out}$ respectively denote the Euclidean distance between pixels $x$ and $y$, the boundary, the inside and outside of the FAZ. Compared to the fixed distance map (DM) suggested in \cite{li2020image} which may produce wrong guidance and fail when the foveal center deviates from the image center and the normal DM which only calculates distance of either foreground or background \cite{tan2018deep,van2019automated}, SDM has two main advantages. It considers the distance transformation information of both foreground and background, which also characterizes the diameter of FAZ and the fovea position. Moreover, since the target area is typically smaller than the background, through respectively normalizing the distance inside and outside FAZ to $[-1, 0]$ and $[0, 1]$, SDM naturally imposes more weights on the interior region and is beneficial for solving the class imbalance issue. As for the boundary regression branch, considering the subjectivity in manual annotations, we generate a Gaussian kernel matrix $\mathcal{G}(\cdot)$ centered at each point $c_n$ on the boundary $\partial G$ and construct a soft label heatmap in the form of $Heatsum$ and treat it as a regression task using mean squared error (MSE) loss instead of treating it as a single pixel boundary segmentation problem:
\begin{equation}
\mathcal{G}(c_n)=\frac{1}{2 \pi \sigma^{2}} e^{-\|x-c_n\|_{2}^{2} / 2 \sigma^{2}}, 
\end{equation}
\begin{equation}
Heatsum(\mathcal{G}(c_1), \mathcal{G}(c_2))=1-(1-\mathcal{G}(c_1))\circ(1-\mathcal{G}(c_2)), 
\end{equation}
\begin{equation}
G_{bd}= Gaussheat(\partial G)=Heatsum(\mathcal{G}(c_1), \mathcal{G}(c_2), ..., \mathcal{G}(c_n)), \forall c_n \in \partial G, 
\end{equation}
where $\circ$ denotes the Hadamard product. Also, $G_{bd}$ is normalized to $[0, 1]$ with values less than 0.001 set to be 0 before normalization. For the segmentation branch $S$, we use Dice loss to evaluate the pixel-wise agreement
between the prediction and the ground truth. Therefore, with trade-off parameters $\lambda_1$, $\lambda_2$, $\lambda_3$, the total objective function of the segmentation model is defined as
\begin{equation}
\begin{split}
\mathcal{L}_{seg} &= \lambda_1\mathcal{L}_{sh} + \lambda_2\mathcal{L}_{bd}+ \lambda_3\mathcal{L}_{sd} \\ 
&= \lambda_1\mathcal{L}_{dice}(\mathcal{S}(p^s_i), G) + \lambda_2\mathcal{L}_{mse}(p^b_i, G_{bd}^{n})+ \lambda_3\mathcal{L}_{mse}(p^d_i, G_{sd}^{n}),
\end{split}
\end{equation}
where $p^s_i$, $p^b_i$, and $p^d_i$ respectively denote the predictions from branches $S$, $B$, and $D$ given an input image $i$, and $\mathcal{S}$ represents the Sigmoid function. $G_{bd}^{n}$ and $G_{sd}^{n}$ are the normalized $G_{bd}$ and $G_{sd}$.

\subsection{Classification Network and Joint Learning Strategy}

In the training phase, the segmentor is first trained via the aforementioned tasks, while the classifier $C$ is initially frozen to avoid instability caused by inaccurate features from the segmentation network. We set a starting flag $\tau$ when the segmentation network almost converges to start joint learning. During joint learning, the multi-level features reconstructed by branches $S$, $B$, and $D$ are sufficiently percepted by the unfrozen classifier in a hierarchical and interpretable way. For each feature concatenation, randomly initialized 1 $\times$ 1 convolutions are adopted for dimensionality reduction. Being consistent with the encoder $E$, we use ResNeSt50 partially (expect the above convolutions) initialized with weights pretrained on ImageNet as the classifier and employ the standard Cross-Entropy loss as $\mathcal{L}_{cl}$. So the final loss of BSDA-Net is defined as (with coefficient $\lambda_0$ to balance the two loss terms):
\begin{equation}
    \mathcal{L}_{joint}=\left\{\begin{array}{cc}
        \mathcal{L}_{seg}, & \text { epoch } \leqslant \tau \\
        \mathcal{L}_{seg}+\lambda_0 \mathcal{L}_{cl}, & \text { epoch }>\tau
        \end{array}\right..
\end{equation}

\section{Experiments and Results}

\subsubsection{Dataset and Preprocessing.}
We evaluate our proposed BSDA-Net framework on three recently released OCTA datasets: OCTA-500, OCTAGON, and Foveal Avascular Zone Image Database (FAZID), the details of which are listed in Table \ref{table1}. In OCTA-500, we only use data from three categories (normal, DR, and AMD) each with a sample size greater than 20. For OCTAGON, the two categories are normal and DR, and FAZID includes normal, diabetic, and myopic. 
For each dataset, to enlarge the sample size and to better evaluate the classification performance, we unify and combine images with different spatial resolutions via resizing and center cropping. 
The final category distributions within the three datasets are listed in Tables {\color{blue} A1} to {\color{blue} A3} (in appendix). 
We adopt data augmentation methods that will not change the spatial resolution and FAZ shape, including random rotation, flipping, contrast adjustment, adding Gaussian noise, and random blurring.

\begin{table}[H]
    \centering
    \renewcommand\arraystretch{1.1}
    \caption{
    Details and preprocessing of the three datasets utilized in our experiments.}
    \label{table1}
\resizebox{11.5cm}{!}{
    \begin{tabular}{c|p{1.6cm}<{\centering}|p{1.9cm}<{\centering}|p{1.6cm}<{\centering}|p{1.6cm}<{\centering}|p{1.7cm}<{\centering}}
    \specialrule{0.13em}{0pt}{0pt}
    Dataset & \multicolumn{2}{c|}{\textbf{OCTA-500\cite{li2020image,li2020ipn}}} & \multicolumn{2}{c|}{\textbf{OCTAGON\cite{diaz2019automatic}}} & \textbf{FAZID\cite{agarwal2020foveal}} \\ 
    \specialrule{0.13em}{0pt}{0pt}
    \begin{tabular}[c]{@{}c@{}} FOV [mm] \end{tabular} & 3  $\times$ 3  & 6 $\times$ 6 & 3 $\times$ 3  & 6 $\times$ 6  & 6 $\times$ 6  \\ \hline
\begin{tabular}[c]{@{}c@{}}Original resolution [px]\end{tabular} & 304 $\times$ 304 & 400 $\times$ 400 & \multicolumn{2}{c|}{320 ×320} & \begin{tabular}[c]{@{}c@{}}420 × 420\end{tabular} \\ \hline
Retinal depth &\multicolumn{2}{c|}{superficial} &\multicolumn{2}{c|}{both superficial and deep}  & superficial \\ \hline
Number & 195 & 169 & 108 & 105 & 304 \\ \hline
Preprocessing & resize & crop + resize & resize & crop & crop \\ \hline
Processed size [px]& \multicolumn{2}{c|}{192 × 192} & \multicolumn{2}{c|}{160 × 160} & 224 × 224 \\ 
\specialrule{0.13em}{0pt}{0pt}

    \end{tabular}}
\end{table}

\subsubsection{Implementation Details.}
All compared models and the proposed BSDA-Net framework are implemented with Pytorch using NVIDIA Tesla A100 GPUs. We adopt ResNeSt50 as the encoder of both the segmentation network and the classifier in this work. We use the Adam optimizer with a learning rate of 1 $\times$ $10^{-4}$ with no learning rate policy for the segmentation network and another separate Adam optimizer with a learning rate of 2 $\times$ $10^{-5}$ for the classifier. Empirically, we set the starting point of the joint learning $\tau$ as 20 and train the network for a total of 200 epochs. Trade-off coefficients $\lambda_0$, $\lambda_1$, $\lambda_2$, $\lambda_3$, and $\sigma$ for soft contour are respectively set to be 1, 3, 1, 1, and 2. For internal validation, we split each dataset into 70\%, 20\%, and 10\% for training, testing, and validation. Five-fold cross-validation is used for fair comparison in all settings. 


\subsubsection{Evaluation of FAZ Segmentation.}
\begin{figure}[H]
    \centering
    \centerline{\includegraphics[width=\textwidth]{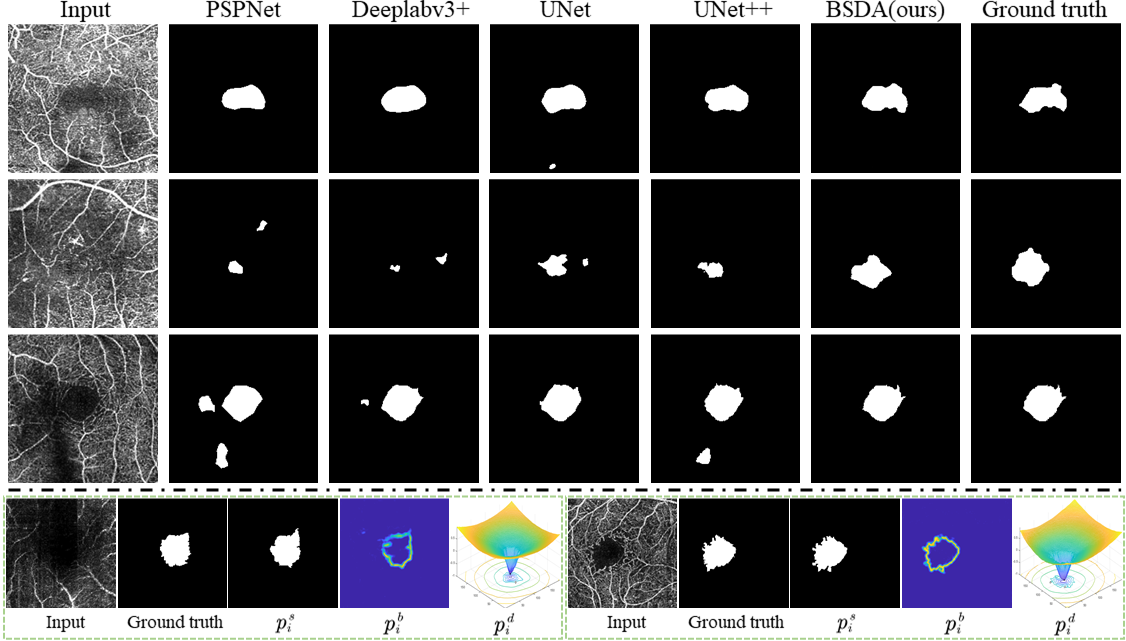}}

    \caption{Visualization results. The upper panel displays segmentation predictions of compared models and the proposed BSDA-Net. The bottom panel shows representative segmentation predictions $p^s_x$, regressed soft boundary $p^b_x$, and reconstructed SDM $p^d_x$. Zoom-in for details.} \label{fig2}
\end{figure}

All methods are evaluated using four metrics, i.e., Dice[\%], Jaccard[\%], Average Symmetric Surface Distance (ASD[px]), and 95\% Hausdorff Distance (95HD[px]) \cite{heimann2009comparison}. Table \ref{table2} tabulates the segmentation results on the three OCTA datasets. We compare BSDA-Net with the baseline ResNeSt-UNet (based on our implementation) and several ablation models (to verify the impact of each auxiliary task), as well as several SOTA segmentation models, e.g., Deeplabv3+, PSPNet for natural image segmentation and UNet, UNet++ for medical image segmentation. By compared with the sixth row, the seventh and eighth rows in Table \ref{table2} indicate that the soft boundary regression constraint and the SDM reconstruction are effective in enhancing the FAZ segmentation performance in terms of every evaluation metric for all three datasets. The last two rows of that table indicate that when jointly learning classification and segmentation, though subsequent results identify the effectiveness of such joint learning strategy for boosting the classification performance, BSDA-Net may not achieve the best segmentation results. In other words, the joint learning framework slightly sacrifices the segmentation performance (without significant difference in any metric of any dataset, \textit{p-}value $>$ 0.05), with a reward of a much greater degree of improvement in the classification performance, as we will show later. Results shown in the penultimate row of Table \ref{table2} can be treated as the upper bound of our FAZ segmentation. 
Fig. \ref{fig2} displays representative segmentation results of the proposed method and four compared models. We also present two sets of intermediate outputs from BSDA-Net, which clearly show the effectiveness of the regressed soft boundary in assessing model's uncertainty and the reconstructed SDM in extracting fovea. 

\begin{table}[htb]
    \centering
    \renewcommand\arraystretch{1}
    \caption{
        Quantitative evaluations of different networks for FAZ segmentation. Joint learning slightly sacrifices the segmentation performance with no significant difference in any metric.}
        \label{table2}
    \resizebox{\textwidth}{!}{
        \begin{tabular}{l|c|c|c|c|c|c|c|c|c|c|c|c}
            
            \specialrule{0.13em}{0pt}{0pt}
            \multicolumn{1}{l|}{\multirow{2}{*}{Method}} & \multicolumn{4}{c|}{\textbf{OCTA-500}} & \multicolumn{4}{c|}{\textbf{OCTAGON}} & \multicolumn{4}{c}{\textbf{FAZID}} \\ \cline{2-13} 
            \multicolumn{1}{c|}{} & Dice$\uparrow$ & Jaccard$\uparrow$ & 95HD$\downarrow$ & ASD$\downarrow$  & Dice$\uparrow$ & Jaccard$\uparrow$  & 95HD$\downarrow$ & ASD$\downarrow$ & Dice$\uparrow$ & Jaccard$\uparrow$  & 95HD$\downarrow$ & ASD$\downarrow$ \\ 
            \specialrule{0.13em}{0pt}{0pt}
            UNet \cite{ronneberger2015u} (w BN)& \begin{tabular}[c]{@{}c@{}}94.51\\ $\pm$ 6.71\end{tabular} & \begin{tabular}[c]{@{}c@{}}90.16\\ $\pm$ 9.23\end{tabular} & \begin{tabular}[c]{@{}c@{}}5.35\\ $\pm$ 9.67\end{tabular} & \begin{tabular}[c]{@{}c@{}}1.02\\ $\pm$ 1.81\end{tabular} & \begin{tabular}[c]{@{}c@{}}87.71\\ $\pm$ 7.09\end{tabular} & \begin{tabular}[c]{@{}c@{}}78.73\\ $\pm$ 9.95\end{tabular} & \begin{tabular}[c]{@{}c@{}}7.63\\ $\pm$ 12.93\end{tabular} & \begin{tabular}[c]{@{}c@{}}2.28\\ $\pm$ 3.01\end{tabular} & \begin{tabular}[c]{@{}c@{}}89.29\\ $\pm$ 8.09\end{tabular} & \begin{tabular}[c]{@{}c@{}}81.40\\ $\pm$ 10.53\end{tabular} & \begin{tabular}[c]{@{}c@{}}7.65\\ $\pm$ 11.48\end{tabular} & \begin{tabular}[c]{@{}c@{}}2.34\\ $\pm$ 2.60\end{tabular} \\

            UNet++ \cite{zhou2019unet++}& \begin{tabular}[c]{@{}c@{}}94.71\\ $\pm$ 6.61\end{tabular} & \begin{tabular}[c]{@{}c@{}}90.53\\ $\pm$ 9.42\end{tabular} & \begin{tabular}[c]{@{}c@{}}5.33\\ $\pm$ 8.61\end{tabular} & \begin{tabular}[c]{@{}c@{}}0.99\\ $\pm$ 1.81\end{tabular} & \begin{tabular}[c]{@{}c@{}}87.86\\ $\pm$ 6.75\end{tabular} & \begin{tabular}[c]{@{}c@{}}78.94\\ $\pm$ 9.90\end{tabular} & \begin{tabular}[c]{@{}c@{}}6.35\\ $\pm$ 9.58\end{tabular} & \begin{tabular}[c]{@{}c@{}}2.05\\ $\pm$ 2.33\end{tabular} & \begin{tabular}[c]{@{}c@{}}88.73\\ $\pm$ 7.94\end{tabular} & \begin{tabular}[c]{@{}c@{}}80.51\\ $\pm$ 10.95\end{tabular} & \begin{tabular}[c]{@{}c@{}}8.26\\ $\pm$ 11.92\end{tabular} & \begin{tabular}[c]{@{}c@{}}2.50\\ $\pm$ 2.81\end{tabular}\\
            PSPNet \cite{zhao2017pyramid}& \begin{tabular}[c]{@{}c@{}}91.48\\ $\pm$ 6.89\end{tabular} & \begin{tabular}[c]{@{}c@{}}84.87\\ $\pm$ 9.29\end{tabular} & \begin{tabular}[c]{@{}c@{}}6.91\\ $\pm$ 11.01\end{tabular} & \begin{tabular}[c]{@{}c@{}}1.63\\ $\pm$ 2.07\end{tabular} & \begin{tabular}[c]{@{}c@{}}86.87\\ $\pm$ 6.69\end{tabular} & \begin{tabular}[c]{@{}c@{}}77.34\\ $\pm$ 9.49\end{tabular} & \begin{tabular}[c]{@{}c@{}}6.89\\ $\pm$ 10.92\end{tabular} & \begin{tabular}[c]{@{}c@{}}2.18\\ $\pm$ 1.98\end{tabular} & \begin{tabular}[c]{@{}c@{}}87.58\\ $\pm$ 9.40\end{tabular} & \begin{tabular}[c]{@{}c@{}}78.91\\ $\pm$ 12.17\end{tabular} & \begin{tabular}[c]{@{}c@{}}10.04\\ $\pm$ 16.94\end{tabular} & \begin{tabular}[c]{@{}c@{}}2.96\\ $\pm$ 3.57\end{tabular} \\
            DeepLabv3+ \cite{chen2018encoder}& \begin{tabular}[c]{@{}c@{}}92.35\\ $\pm$ 7.33\end{tabular} & \begin{tabular}[c]{@{}c@{}}86.44\\ $\pm$ 9.71\end{tabular} & \begin{tabular}[c]{@{}c@{}}6.06\\ $\pm$ 7.01\end{tabular} & \begin{tabular}[c]{@{}c@{}}1.34\\ $\pm$ 1.49\end{tabular} & \begin{tabular}[c]{@{}c@{}}87.28\\ $\pm$ 6.82\end{tabular} & \begin{tabular}[c]{@{}c@{}}78.00\\ $\pm$ 9.50\end{tabular} & \begin{tabular}[c]{@{}c@{}}6.65\\ $\pm$ 10.39\end{tabular} & \begin{tabular}[c]{@{}c@{}}2.17\\ $\pm$ 2.38\end{tabular} & \begin{tabular}[c]{@{}c@{}}88.22\\$\pm$ 8.98\end{tabular} & \begin{tabular}[c]{@{}c@{}}79.82\\ $\pm$ 11.57\end{tabular} & \begin{tabular}[c]{@{}c@{}}8.02\\ $\pm$ 13.57\end{tabular} & \begin{tabular}[c]{@{}c@{}}2.74\\ $\pm$ 4.12\end{tabular}  \\ 
            \specialrule{0.1em}{0pt}{0pt}
            Baseline ($E$ + $S$) & \begin{tabular}[c]{@{}c@{}}95.24\\ $\pm$ 7.00\end{tabular} & \begin{tabular}[c]{@{}c@{}}91.49\\ $\pm$ 9.05\end{tabular} & \begin{tabular}[c]{@{}c@{}}4.52\\ $\pm$ 7.20\end{tabular} & \begin{tabular}[c]{@{}c@{}}0.89\\ $\pm$ 2.25\end{tabular} & \begin{tabular}[c]{@{}c@{}}87.73\\ $\pm$ 7.82\end{tabular} & \begin{tabular}[c]{@{}c@{}}78.84\\ $\pm$ 10.24\end{tabular} & \begin{tabular}[c]{@{}c@{}}6.95\\ $\pm$ 11.61\end{tabular} & \begin{tabular}[c]{@{}c@{}}2.19\\ $\pm$ 2.81\end{tabular} & \begin{tabular}[c]{@{}c@{}}89.89\\$\pm$ 7.16\end{tabular} & \begin{tabular}[c]{@{}c@{}}82.27\\ $\pm$ 10.02\end{tabular} & \begin{tabular}[c]{@{}c@{}}7.05\\ $\pm$ 10.36\end{tabular} & \begin{tabular}[c]{@{}c@{}}2.20\\ $\pm$ 2.57\end{tabular}  \\
            BSDA (w/o $D$, $C$) & \begin{tabular}[c]{@{}c@{}}95.90\\$\pm$ 4.20\end{tabular} & \begin{tabular}[c]{@{}c@{}}92.40\\ $\pm$ 6.71\end{tabular} & \begin{tabular}[c]{@{}c@{}}4.37\\ $\pm$ 7.26\end{tabular} & \begin{tabular}[c]{@{}c@{}}0.76\\ $\pm$ 1.48\end{tabular} & \begin{tabular}[c]{@{}c@{}}88.30\\$\pm$ 6.32\end{tabular} & \begin{tabular}[c]{@{}c@{}}79.56\\ $\pm$ 9.04\end{tabular} & \begin{tabular}[c]{@{}c@{}}6.23\\ $\pm$ 9.81\end{tabular} & \begin{tabular}[c]{@{}c@{}}2.00\\ $\pm$ 2.02\end{tabular} & \begin{tabular}[c]{@{}c@{}}90.77\\$\pm$ 5.87\end{tabular} & \begin{tabular}[c]{@{}c@{}}83.57\\ $\pm$ 8.63\end{tabular} & \begin{tabular}[c]{@{}c@{}}6.52\\ $\pm$ 10.15\end{tabular} & \begin{tabular}[c]{@{}c@{}}1.98\\ $\pm$ 2.31\end{tabular}\\
            BSDA (w/o $B$, $C$) & \begin{tabular}[c]{@{}c@{}}95.84\\$\pm$ 4.40\end{tabular} & \begin{tabular}[c]{@{}c@{}}92.30\\ $\pm$ 6.92\end{tabular} & \begin{tabular}[c]{@{}c@{}}4.08\\ $\pm$ 5.81\end{tabular} & \begin{tabular}[c]{@{}c@{}}0.72\\ $\pm$ 1.20\end{tabular} & \begin{tabular}[c]{@{}c@{}}88.14\\$\pm$ 6.45\end{tabular} & \begin{tabular}[c]{@{}c@{}}79.31\\ $\pm$ 9.10\end{tabular} & \begin{tabular}[c]{@{}c@{}}5.94\\ $\pm$ 9.63\end{tabular} & \begin{tabular}[c]{@{}c@{}}1.97\\ $\pm$ 2.18\end{tabular} & \begin{tabular}[c]{@{}c@{}}90.60\\$\pm$ 7.12\end{tabular} & \begin{tabular}[c]{@{}c@{}}83.43\\ $\pm$ 9.57\end{tabular} & \begin{tabular}[c]{@{}c@{}}6.19\\ $\pm$ 7.42\end{tabular} & \begin{tabular}[c]{@{}c@{}}1.90\\ $\pm$ 2.26\end{tabular}\\

            BSDA (w/o $C$) & \begin{tabular}[c]{@{}c@{}}\textbf{96.21}\\\textbf{$\pm$ 3.78}\end{tabular} & \begin{tabular}[c]{@{}c@{}}\textbf{92.92}\\ \textbf{$\pm$ 6.14}\end{tabular} & \begin{tabular}[c]{@{}c@{}}\textbf{3.61}\\ \textbf{$\pm$ 5.68}\end{tabular} & \begin{tabular}[c]{@{}c@{}}\textbf{0.63}\\ \textbf{$\pm$ 1.23}\end{tabular} & \begin{tabular}[c]{@{}c@{}}\textbf{88.64}\\\textbf{$\pm$ 5.89}\end{tabular} & \begin{tabular}[c]{@{}c@{}}\textbf{80.00}\\ \textbf{$\pm$ 8.66}\end{tabular} & \begin{tabular}[c]{@{}c@{}}\textbf{5.43}\\ \textbf{$\pm$ 6.86}\end{tabular} & \begin{tabular}[c]{@{}c@{}}\textbf{1.80}\\ \textbf{$\pm$ 1.26}\end{tabular} & \begin{tabular}[c]{@{}c@{}}\textbf{91.03}\\\textbf{$\pm$ 5.13}\end{tabular} & \begin{tabular}[c]{@{}c@{}}\textbf{83.91}\\ \textbf{$\pm$ 7.92}\end{tabular} & \begin{tabular}[c]{@{}c@{}}5.85\\ $\pm$ 6.05\end{tabular} & \begin{tabular}[c]{@{}c@{}}\textbf{1.81}\\ \textbf{$\pm$ 1.23}\end{tabular} \\
            BSDA (ours) & \begin{tabular}[c]{@{}c@{}}96.07\\$\pm$ 4.28\end{tabular} & \begin{tabular}[c]{@{}c@{}}92.72\\ $\pm$ 6.82\end{tabular} & \begin{tabular}[c]{@{}c@{}}3.90\\ $\pm$ 6.03\end{tabular} & \begin{tabular}[c]{@{}c@{}}0.68\\ $\pm$ 1.28\end{tabular} & \begin{tabular}[c]{@{}c@{}}88.37\\$\pm$ 6.03\end{tabular} & \begin{tabular}[c]{@{}c@{}}79.64\\ $\pm$ 8.75\end{tabular} & \begin{tabular}[c]{@{}c@{}}5.79\\ $\pm$ 8.95\end{tabular} & \begin{tabular}[c]{@{}c@{}}1.92\\ $\pm$ 1.96\end{tabular} & \begin{tabular}[c]{@{}c@{}}90.98\\$\pm$ 5.19\end{tabular} & \begin{tabular}[c]{@{}c@{}}83.84\\ $\pm$ 8.03\end{tabular} & \begin{tabular}[c]{@{}c@{}}\textbf{5.67}\\ \textbf{$\pm$ 5.53}\end{tabular} & \begin{tabular}[c]{@{}c@{}}1.82\\ $\pm$ 1.23\end{tabular} \\ 
            \specialrule{0.13em}{0pt}{0pt}
            \end{tabular}}
\end{table}

\begin{table}[htb]
    \centering
    \renewcommand\arraystretch{0.92}
    \caption{
        Classification performance of all methods on the three OCTA datasets. }

        \label{table3}
    \resizebox{10cm}{!}{
        \begin{tabular}{l|c|p{0.95cm}<{\centering}p{0.95cm}<{\centering}|p{0.95cm}<{\centering}p{0.95cm}<{\centering}|p{0.95cm}<{\centering}p{0.95cm}<{\centering}}
            \specialrule{0.13em}{0pt}{0pt}
            \multirow{2}{*}{Method} & \multirow{2}{*}{\begin{tabular}[c]{@{}c@{}}Pretrained\\ (ImageNet)\end{tabular}} & \multicolumn{2}{c|}{\textbf{OCTA-500}} & \multicolumn{2}{c|}{\textbf{OCTAGON}} & \multicolumn{2}{c}{\textbf{FAZID}} \\ \cline{3-8} 
             &  & Acc & Kappa & Acc & Kappa & Acc & Kappa \\ 
             \specialrule{0.13em}{0pt}{0pt}
            \multirow{2}{*}{VGG16 \cite{andreeva2020dr,le2020transfer,simonyan2014very}} & - & 84.62 & 64.23 & 95.31 & 89.36 & 66.78 & 49.35 \\
             & $\checkmark$ & 86.53 & 72.40 & 96.24 & 91.55 & 70.72 & 55.63 \\ 
            \multirow{2}{*}{ResNet50 \cite{he2016deep}} & - & 81.59 & 60.01 & 90.61 & 78.23 & 70.72 & 55.68 \\
             & $\checkmark$ & 89.84 & 77.88 & \textbf{97.65} & \textbf{94.58} & 74.67 & 61.70 \\ 
            \multirow{2}{*}{ResNeXt50 \cite{xie2017aggregated}} & - & 82.97 & 62.16 & 87.79 & 71.92 & 70.72 & 55.78 \\
             & $\checkmark$ & 90.11 & 78.38 & 96.24 & 91.29 & 72.37 & 58.10 \\ 
            \multirow{2}{*}{ResNeSt50 \cite{zhang2020resnest}} & - & 89.01 & 76.62 & 95.77 & 90.32 & \textbf{75.00} & \textbf{62.08} \\
             & $\checkmark$ & \textbf{90.93} & \textbf{80.32} & 96.71 & 92.53 & \textbf{75.00} & \textbf{62.08} \\ 
             \specialrule{0.1em}{0pt}{0pt}
            YNet (ResNeSt50)\cite{mehta2018net} & $\checkmark$ & 91.76 & 82.21 & 96.71 & 92.53 & 74.34 & 61.24 \\ 
            BSDA (w/o $B$, $D$) & $\checkmark$ & 92.03 & 82.67 & 97.65 & 94.58 & 78.62 & 67.81 \\ 
            BSDA (ours) & $\checkmark$ & \textbf{94.23} & \textbf{87.68} & \textbf{99.53} & \textbf{98.92} & \textbf{82.57} & \textbf{73.67} \\ 
            \specialrule{0.13em}{0pt}{0pt}
            \end{tabular}}
\end{table}

\subsubsection{Evaluation of OCTA Classification.}
As for classification, accuracy and Cohen’s kappa are adopted to evaluate all models. The compared methods include VGG16 (employed in previously proposed OCTA classification works), ResNet50, ResNeXt50, and the SOTA ResNeSt50 (with and without ImageNet pretrained). Moreover, we compare with the YNet \cite{mehta2018net} of a shared-encoder structure for joint classification and segmentation (based on our reimplementation using ResNeSt50 as the encoder), and the ablation structure without branches $B$ and $D$. The quantitative results are shown in Table \ref{table3}, which demonstrates that our model achieves the best performance and identifies the effectiveness of FAZ features and the proposed hierarchical feature perception strategy in boosting deep classification networks. Comparing results listed in the last two rows of Table \ref{table3}, there is no doubt that our proposed joint learning strategy benefits the classification significantly, with only a very mild decrease in the segmentation performance (Table \ref{table2}). In addition, we display detailed classification reports, including the precision, recall, F1-score of each disease and the macro avg, weight avg of each dataset, in Tables {\color{blue} A1} to {\color{blue} A3} (from our appendix) for the three datasets. These results establish new SOTA classification baselines for OCTA images. 
Across all three datasets, BSDA-Net is found to yield the best classification results on DR. Even in FAZID, samples misclassified in the diabetic category are mainly diabetic non-retinopathy samples. The classification performance on myopia in the FAZID dataset is relatively inferior (79.68\%), which partially agrees with existing evidence that FAZ features are relatively indistinguishable in low-moderate myopia \cite{agarwal2020foveal,balaji2020comparison}.


\section{Conclusion}

This paper presents a novel hierarchical and multi-level boundary shape and distance aware joint learning (BSDA-Net) framework for FAZ segmentation and diagnostic classification. 
Specifically, by constructing a boundary heatmap regression branch and a SDM reconstruction branch, we essentially propose a soft contour and directional signed distance aware segmentation loss, which is found to predict more accurate FAZ boundaries and suppress outliers. 
We also design a hierarchical and interpretable joint learning strategy to fuse FAZ features with those from the classifier. 
Extensive experiments on three publicly-accessible OCTA datasets show that our BSDA-Net achieves significantly better performance than SOTA methods on both segmentation and classification.
Collectively, our results demonstrate the potential of OCTA for automated ophthalmological and systemic disease assessments and the effectiveness of FAZ features in boosting deep learning classifiers' performance.

\newpage
\section*{Appendix}

\appendix
\renewcommand\thetable{\Alph{section}\arabic{table}}
\setcounter{table}{0}
\renewcommand{\thetable}{A\arabic{table}}
\renewcommand\thefigure{\Alph{section}\arabic{figure}}  
\setcounter{figure}{0}
\renewcommand{\thefigure}{A\arabic{figure}}


\vspace{-0.6cm}

\begin{table}[H]
    \centering
    \caption{BSDA-Net's classification results on OCTA-500.}
    \renewcommand\arraystretch{1.05}
    \begin{tabular}{p{1.9cm}<{\centering}|p{1.4cm}<{\centering}|p{1.4cm}<{\centering}|p{1.4cm}<{\centering}|p{1.4cm}<{\centering}}
        \specialrule{0.13em}{0pt}{0pt}
        & Precision & Recall & F1-score & Support \\ 
        \specialrule{0.13em}{0pt}{0pt}
        Normal & 95.31 & 97.21 & 96.25 & 251 \\ 
        DR & 94.92 & 87.50 & 91.06 & 64 \\ 
        AMD & 87.76 & 87.76 & 87.76 & 49 \\ 
        \specialrule{0.13em}{0pt}{0pt}
        \textbf{Macro avg} & 92.66 & 90.82 & 91.69 & \multirow{3}{*}{364}  \\ 
        \textbf{Weight avg} & 94.23 & 94.23 & 94.20 &  \\ 
        \Xcline{1-4}{1.05pt}
        \textbf{Accuracy} & 94.23  & \textbf{Kappa} & 87.68 \\ 
        
        \specialrule{0.13em}{0pt}{0pt}
        \end{tabular}
        \label{table_1} 
\end{table}

\vspace{-0.6cm}

\begin{table}[H]
    \centering
    \renewcommand\arraystretch{1.05}
    \caption{BSDA-Net's classification results on OCTAGON.}
    \begin{tabular}{p{1.9cm}<{\centering}|p{1.4cm}<{\centering}|p{1.4cm}<{\centering}|p{1.4cm}<{\centering}|p{1.4cm}<{\centering}}
        \specialrule{0.13em}{0pt}{0pt}
        & Precision & Recall & F1-score & Support \\ 
        \specialrule{0.13em}{0pt}{0pt}
        Normal & 99.31 & 100 & 99.65 & 144 \\ 
        DR & 100 & 98.55 & 99.27 & 69 \\ 
        \specialrule{0.13em}{0pt}{0pt}
        
        \textbf{Macro avg}& 99.66 & 99.28 & 99.46 & \multirow{3}{*}{213}  \\ 
        \textbf{Weight avg} & 99.53 & 99.53 & 99.53 &  \\ 
        \Xcline{1-4}{1.05pt}
        \textbf{Accuracy} & 99.53 & \textbf{Kappa} & 98.92 \\
        \specialrule{0.13em}{0pt}{0pt}
        \end{tabular} 
      \label{table_2} 
\end{table}

\vspace{-0.6cm}

\begin{table}[H]
    \centering
      \renewcommand\arraystretch{1.05}
      \caption{
        BSDA-Net's classification results on FAZID.}\label{table_3}
    \begin{tabular}{p{1.9cm}<{\centering}|p{1.4cm}<{\centering}|p{1.4cm}<{\centering}|p{1.4cm}<{\centering}|p{1.4cm}<{\centering}}
        \specialrule{0.13em}{0pt}{0pt}
        & Precision & Recall & F1-score & Support \\ 
        \specialrule{0.13em}{0pt}{0pt}
        Normal & 74.39 & 69.32 & 71.77 & 88 \\ 
        Diabetic & 92.93 & 85.98 & 89.32 & 107 \\  
        Myopic & 79.68 & 89.91 & 84.48 & 109 \\ 
        \specialrule{0.13em}{0pt}{0pt}
        \textbf{Macro avg} & 82.33 & 81.74 & 81.86 & \multirow{3}{*}{304}  \\ 
        \textbf{Weight avg} & 82.81 & 82.57 & 82.50 &  \\ 
        \Xcline{1-4}{1.05pt}
        \textbf{Accuracy} & 82.57  & \textbf{Kappa} & 73.67 \\ 
        \specialrule{0.13em}{0pt}{0pt}
        \end{tabular}
\end{table}

\vspace{-0.6cm}

\begin{figure}[htb]
    \centering
    \centerline{\includegraphics[width=\textwidth]{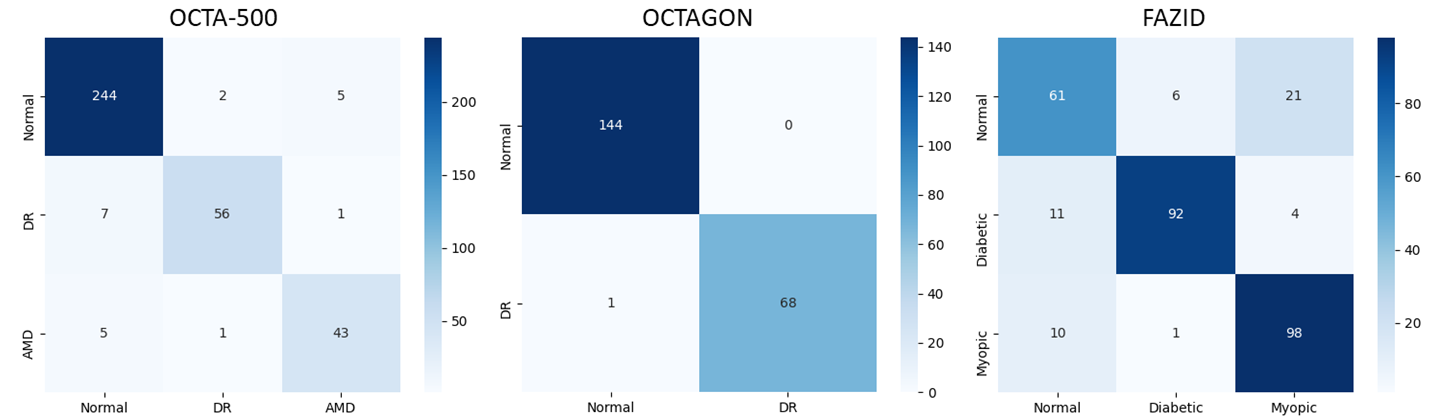}}
    \caption{Classification confusion matrixes of BSDA-Net on the three datasets.} \label{fig_a0}
\end{figure}

\vspace{-0.6cm}

\begin{figure}[htbp]
    \centering
    \centerline{\includegraphics[width=\textwidth]{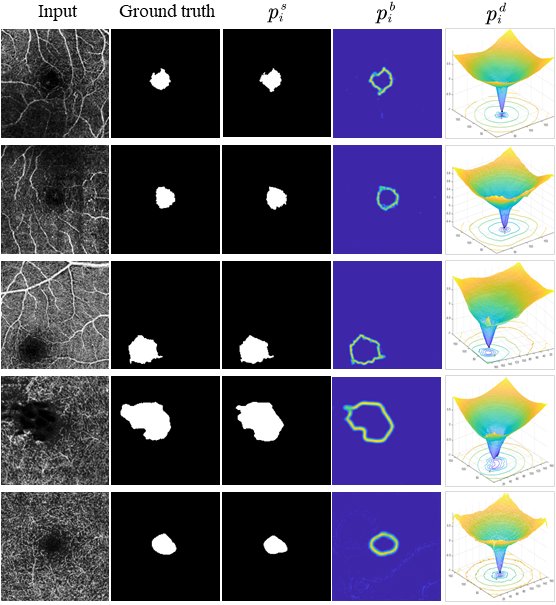}}
    \caption{Visualization results. Representative segmentation predictions $p^s_i$, regressed soft boundary $p^b_i$, and reconstructed SDM $p^d_i$ from BSDA-Net. Zoom-in for details.} \label{fig_a1}
\end{figure}

\end{document}